\documentclass[11pt]{article}
\usepackage{mb,epsfig}

\def\Journal#1#2#3#4{{#1} {\bf #2}, #3 (#4)}

\def\NPB{{\em Nucl. Phys.} B}
\def\PLB{{\em Phys. Lett.}  B}
\def\PRL{\em Phys. Rev. Lett.}
\def\PRD{{\em Phys. Rev.} D}

\def\be{\begin{equation}}
\def\ee{\end{equation}}
\def\bea{\begin{eqnarray}}
\def\eea{\end{eqnarray}}

\begin{document}

\begin{flushright}
        CERN-TH/98-202\\
        hep-ph/9806429 \hspace*{0.1cm}\\ 
        June 1998 \mbox{\hspace*{0.6cm}}\\
\end{flushright}

\vspace*{2cm}
\title{NEW RESULTS ON HEAVY QUARKS NEAR THRESHOLD$\,$\footnote{
Talk presented at the XXXIIIrd Rencontres de Moriond 
`Electroweak Interactions and Unified Theories', 14-21 March 1998, 
Les Arcs, France. Compared to the presentation at the meeting, 
this write-up has been updated to account for results published 
after the date of the conference.}}

\author{M. BENEKE}

\address{Theory Division, CERN, CH-1211 Geneva, Switzerland}

\maketitle\abstracts{
We review in brief the threshold expansion, a method to perform 
the expansion of Feynman integrals near the heavy quark-antiquark 
threshold, and its relation to the construction of two effective 
theories, non-relativistic QCD (NRQCD) and potential-NRQCD. We then 
summarize recent next-to-next-to-leading order results on the 
decay $J/\Psi\to l^+ l^-$, the bottom quark mass from $\Upsilon$ 
sum rules and the top-anti-top production cross section near threshold 
in $e^+ e^-$ collisions.}

\section{Motivation}

In this talk we discuss systems of the form $Q\bar{Q}+X$, where 
$Q$ is a heavy quark of mass $m$ and $X$ a collection of massless 
particles, in the kinematic region, where the total invariant mass 
$q^2$ of the system is close to $4 m^2$. The physics of such systems is 
characterized by the fact that in the 
reference frame with $\vec{q}=0$ the heavy quark velocities are 
small, i.e.\ $v\ll 1$.

The threshold region is evidently sensitive to the mass of the quark 
and physical quantities that probe the threshold region can therefore be 
used to determine the bottom and top quark mass. In case of bottom 
quarks the threshold region is populated by narrow Upsilon resonances 
and the cross section $e^+ e^-\to b\bar{b}+X$ cannot be predicted 
locally. However, dispersion relations equate an average over 
Upsilon resonances and the continuum to the derivatives of $b$-quark 
current correlation function at $q^2=0$, which can be calculated in 
perturbation theory~\citer{NSVZ1,VZ87}. In case of top quarks the 
rapid decay  
of the top prohibits the formation of long-lived resonances~\cite{B86}. For 
$m_t=175\,$GeV a resonance-like structure remains visible in the energy 
dependence of the cross section $e^+ e^-\to t\bar{t}+X$~\cite{KK,SP91}. 
Perturbation theory can still be applied in the vicinity of the 
`pseudo-resonance', but not arbitrarily close to threshold. The 
shape of the $t\bar{t}$ cross section near threshold, to be measured at 
a Next Linear Collider, is believed to provide us with the most 
accurate determination of the top quark mass, if the strong 
interaction corrections are indeed well understood.

In addition to quark masses as parameters of the standard model, there 
are more intrinsically QCD-related problems involving heavy quarks 
near threshold. Heavy quarkonia are non-relativistic and their 
production and decay can be treated as an expansion~\cite{BBL} in $v$. 
One can also consider heavy quark production in hadron-hadron collisions. 
In this case one is mainly concerned with the resummation of logarithms 
of $v$ that arise as a consequence of soft and collinear gluon emission, 
predominantly from the massless initial state particles. This last 
application will not be discussed here (see, for instance,  
Refs.~\cite{KS96,BCMN98}). Incidentally, we note that at 
Tevatron centre-of-mass energies the average velocity squared of top 
quarks is $\langle v^2\rangle \sim 1/2$ and the non-relativistic 
dynamics discussed subsequently is probably not important there.

A non-relativistic system involves more than one momentum scale, related 
to the small parameter $v$. The first part of the problem consists of 
organizing the calculation in a systematic expansion in $v$ and has been 
addressed in several papers recently~\citer{LAB,Griesshammer}, 
especially in the context of dimensional regularization. These works 
clarify the definition of non-relativistic QCD (NRQCD)~\citer{CL86,LMM} 
with a matching prescription based on dimensional 
regularization and show that one can proceed to a second effective 
theory by integrating out more scales. This development is summarized in 
the first part of the talk.

The second part summarizes recent next-to-next-to-leading order (NNLO)  
results on the phenomenological applications mentioned above. `NNLO' 
in this context refers either to fixed-order calculations in 
$\alpha_s$~\citer{Pet97,CM98}, required for matching effective theories, 
or the resummation of Coulomb-enhanced corrections to all 
orders~\citer{PY97,BSS98II}, required for the solution of the low-energy 
problem.

\section{Complications near the threshold}

The need for resummation arises, because even when $\alpha_s$ is small  
the effective interaction between a heavy quark and anti-quark 
becomes strong at small relative velocity. The exchange of a Coulomb 
gluon -- the $00$-component of the propagator for a gluon with momentum 
of order $m v$ and energy of order $m v^2$ -- leads to an effective 
coupling of order $\pi\alpha_s(m v)/v$. Resummation leads to weakly 
coupled bound states ($E_{\rm bind} \sim m v^2 \ll m$) --  the Coulomb 
bound states analogous to hydrogen and positronium. In first 
approximation the physics is indeed exactly as in 
QED. However, due to massless quarks and gluons, the coupling 
runs below the scale $m$ in QCD. This leads to complications, 
and differences in 
the power counting, when $m v^2\sim \Lambda_{\rm QCD}$.

Defining $v=(1-4 m^2/q^2)^{1/2}$, the cross section 
$e^+ e^-\to Q\bar{Q}+X$ can be expanded in a double series in $\alpha_s$ 
and $v$:
\begin{equation}
\label{exp}
R_{Q\bar{Q}} = v \sum_{k=0}\,\sum_{l=-k} c_{kl}\,\alpha_s^k\,v^l
\times\,\mbox{logs of}\,\,v.
\end{equation}
A NNLO calculation in the kinematic region where $\alpha_s\sim v$ 
has to account for all terms with $k+l\leq 2$. Powers of $v$ arise 
from ratios of momentum scales. We have to disentangle the contributions 
from the different scales in order to be sure that for a high-order 
loop graph, which cannot be calculated exactly, we have taken 
into account all terms with $k+l\leq 2$.

Eventually we will be led to calculating diagrams with Coulomb Green 
functions rather than free Green functions. There are, however, 
two 2-loop calculations in conventional perturbation theory that 
enter at NNLO: (a) The 2-loop correction to the Coulomb potential, 
i.e. $a_2$ in 
\begin{equation}
V(\vec{q}\,) = -\frac{4\pi\alpha_s}{\vec{q}^{\,2}}\,
\sum_k a_k \alpha_s^k.
\end{equation}
It has been calculated in Ref.~\cite{Pet97}. (b) The 2-loop correction 
to the matching of the vector heavy quark current in QCD to its 
non-relativistic analogue, i.e. $c_2$ in 
\begin{equation}
\label{current}
\bar{Q}\gamma_i Q = \left(\sum_k c_k \alpha_s^k\right)
\,\psi^\dagger\sigma_i\chi + \ldots,
\end{equation}
where $\psi$ and $\chi$ are 2-spinors. Only the spatial component is 
needed at NNLO. The coefficient $c_2$ has been calculated 
in Refs.~\cite{BSS98,CM98}. Note that the series expansion of the 
Coulomb potential is uniquely 
specified (at least to order $\alpha_s^2$) by choosing a scheme 
for coupling renormalization. But the non-relativistic current is 
not conserved and one has to choose a factorization scheme to define 
the $c_k$. In Refs.~\cite{BSS98,CM98} the $\overline{\rm MS}$ scheme 
is used.

\section{Scales and their Separation}

We now discuss how to construct the expansion (\ref{exp}), first 
at a given loop order $k$, then including resummation.

\subsection{Threshold expansion of Feynman integrals}

Take a loop integral that contributes to the heavy quark cross section. 
To be precise, we consider the two-point function of the 
current (\ref{current}) and take the imaginary part at the 
end. This avoids calculating the threshold expansion of phase space 
integrals. The threshold expansion \cite{BS} is a method to 
calculate the expansion in $v$ without calculating the full 
integral. Inspection of the denominators of the Feynman integrand 
shows that there are four essential regions of loop momentum:
\begin{eqnarray}
\label{terminology}
\mbox{{\em hard} (h):} && l_0\sim m,\,\,\vec{l}\sim m, \nonumber\\
\mbox{{\em soft} (s):} && l_0\sim m v,\,\,\vec{l}\sim m v,\nonumber\\
\mbox{{\em potential} (p):} && l_0\sim m v^2,\,\,\vec{l}\sim m v,\\ 
\mbox{{\em ultrasoft} (us):} &&  l_0\sim m v^2,\,\,\vec{l}\sim m v^2.
\nonumber
\end{eqnarray}
The threshold expansion is constructed by writing the diagram as 
a sum of terms that follow by dividing each loop momentum integration 
into these four regions. More precisely, one has to account for the 
possibility that for example $l_1$ and $l_2$ are hard, but their sum 
is not. That is, one should sum over one-particle-irreducible 
subgraphs. Note that the division is done implicitly, through the 
expansions of the propagators. No explicit cut-offs are needed. 
The soft region has often been omitted in the discussion of 
non-relativistic dynamics, probably because it gives rise only 
to instantaneous potentials as we discuss below. However, it also 
gives rise to the (standard) evolution of the strong coupling 
between the scales $m$ and $m v$, as can be seen from the fact 
that a light quark loop inserted into a potential gluon line can 
only be hard or soft.

Then, in each term, the propagators can be simplified. The propagator 
of a heavy quark with momentum $(q/2+l_0,\vec{p}+\vec{l}\,)$ is 
\begin{equation}
\frac{1}{l_0^2-\vec{l}^2-2\vec{p}\cdot\vec{l}-\vec{p}^{\,2}+q l_0+
y+i\epsilon},
\end{equation}
and we will assume that $\vec{p}$ scales as $m v$ and 
$y=q^2/4-m^2$ as $m v^2$. When $l$ is hard, 
we expand the terms involving $\vec{p}$ and $y$ 
and the leading term in the expansion scales 
as $v^0$. When $l$ is soft, the term $q l_0$ is largest and the remaining 
ones are expanded. The propagator becomes static and scales as $v^{-1}$. 
When $l$ is potential, the propagator takes its standard non-relativistic 
form after expansion of $l_0^2$ and scales as $v^{-2}$. Massive 
particles can never be ultrasoft. The gluon propagator takes its 
usual form, when the gluon line is soft and ultrasoft and scales 
as $v^{-2}$ and $v^{-4}$, respectively. If the gluon momentum is potential, 
one can expand $l_0^2$ and the interaction becomes instantaneous. 
If we add the scaling rules for the loop integration measure, 
$d^4l\sim 1\,\mbox{(h)},\,v^4\,\mbox{(s)},\,v^5\,\mbox{(p)},\,v^8\mbox{(us)}$, 
we can immediately estimate the size of the leading term from a given 
region. Because all terms that are small in a given region are 
expanded in the Feynman integrand, each term in the resulting sum 
contributes only to a single power of $v$ (multiplied, in general, by 
powers of logarithms of $v$). It is important that the power 
behaviour can be determined before calculating. It is also important 
that the integrals that appear in each term of the expansion are 
much easier to calculate than the original integral, because they are 
homogeneous in $v$, that is, contain only a single scale. 
In particular, any term 
in the expansion of massive 2-loop 3-point 
integrals is calculable \cite{BS,CM98}. 

The diagrammatic rules for the expansion of propagators (and vertices) 
can be considered as following from an effective Lagrangian. Because 
different loop momentum regions do not overlap, one can introduce 
separate fields for hard, potential, etc. quarks and gluons. (For 
the coupling of potential quarks to ultrasoft gluons, the threshold 
expansion entails the multipole expansion.) This has been done in 
Ref.~\cite{Griesshammer} and, in part, in Refs.~\citer{LM,LS}. One 
can then think of integrating out first the hard fields, then  
soft fields and potential gluons.

One can use Coulomb gauge or a covariant gauge. (In covariant gauge 
the ghost fields are treated like other massless fields. The can be 
hard, soft, potential and ultrasoft.) The hard region is 
very inconvenient to calculate in Coulomb gauge. But Coulomb gauge 
makes gauge cancellations manifest that occur in the coupling of soft 
and ultrasoft gluons to heavy quarks. In fact, one can 
use different gauges for different regions with one exception: The 
distinction between soft and potential gluons is not gauge-invariant. 
This should be no surprise. When one integrates out soft gluons alone, 
one has to compute graphs with external potential gluons. But 
potential gluons are off their mass shell. On the other hand, when 
one integrates out soft and potential gluons together, only 
on-shell graphs have to be considered. 

\subsection{NRQCD}

First we integrate out the hard region. This means that we assume 
that all loop momenta are hard and the external momenta soft, 
potential or ultrasoft. That is, we Taylor-expand the integrand in 
the external momenta of the graph. The result is 
obviously polynomial in the external 
momenta and can therefore be written as a local operator. This 
yields the non-relativistic QCD (NRQCD) Lagrangian \citer{CL86,LMM}
\begin{eqnarray}
\label{nrqcd}
{\cal L}_{\rm NRQCD} &=& \psi^\dagger \left(i D^0+\frac{\vec{D}^2}{2 m}
\right)\psi + \frac{1}{8 m^3}\,\psi^\dagger\vec{D}^4\psi-
\frac{g_s}{2 m}\,\psi^\dagger\vec{\sigma}\cdot \vec{B}\psi
\nonumber\\
&&\hspace*{-1.5cm}
-\,\frac{g_s}{8 m^2}\,\psi^\dagger\left(\vec{D}\cdot\vec{E}-
\vec{E}\cdot\vec{D}\right)\psi- 
\frac{i g_s}{8 m^2}\,\psi^\dagger\vec{\sigma}\cdot\left(
\vec{D}\times\vec{E}-\vec{E}\times\vec{D}\right)\psi 
\nonumber\\[0.4cm]
&&\hspace*{-1.5cm}
+\,\,\mbox{antiquark terms}\,+ {\cal L}_{\rm light}
\end{eqnarray}
and an expansion of QCD operators (such as the vector current) in 
terms of non-relativistic fields. We have written down only those terms 
of the Lagrangian that are needed for the NNLO calculations described 
below, provided one treats the difference between 
$\alpha_s(m)$ and $\alpha_s(m v^2)$ as small. Otherwise 
one should use the renormalization group to sum up the leading 
logarithms sensitive to the scale $m v^2$ in the coefficient 
functions. 

Note that the threshold expansion provides a matching prescription, 
in which one does not have to calculate NRQCD diagrams explicitly. 
The QCD diagram is already broken up into the contributions from 
different scales and the hard regions are exactly the relativistic 
effects which are contained in the coefficient functions of the 
effective Lagrangian. The matching prescription is very simple. Despite 
the fact that the QCD diagrams are divergent, when the relative 
momentum of the heavy quarks goes to zero, the matching 
coefficients are given by the Taylor expansion around zero relative 
momentum. For the single-heavy quark sector, this matching 
prescription coincides with that of Ref.~\cite{Ma97}.

Note also that perturbative calculations with the NRQCD Lagrangian 
have up to now been mainly done with a cutoff regularization, either 
in QED or lattice QCD, while we assume dimensional regularization 
here. The change is not quite trivial. For, if we calculated 
a NRQCD integral according to the Feynman rules of its Lagrangian, 
we would obtain an incorrect result, because dimensional regularization 
treats the UV cutoff of NRQCD as larger than $m$. We can use 
dimensional regularization provided we expand the integrand before 
integration. The precise prescription is again specified by the 
threshold expansion. The NRQCD integral is given by expanding the 
integrand according to the rules for the soft, potential and 
ultrasoft region. Indeed, since the hard region is accounted for 
by the coefficient functions, this reproduces the original QCD 
diagram.

The interaction terms in NRQCD do not have a definite scaling 
in $v$, because a gluon field can be soft, potential or ultrasoft, and a 
quark field can be soft or potential and different scaling rules 
apply to each of these. 

\subsection{Potential NRQCD}

If the scale $m v$ is already non-perturbative, we stop with 
NRQCD. If it is not, the threshold expansion suggests that one 
can integrate out the soft region together with potential gluons 
to arrive at another effective Lagrangian. We call the effective 
theory potential NRQCD (\mbox{PNRQCD}), following Refs.~\cite{PS,PS2}, 
where the corresponding tree-level Lagrangian was considered first. 
We cannot integrate out potential quarks, because the 
Green functions relevant to our applications have external 
potential quark lines. 

To integrate out the soft contributions we consider (NRQCD) 
graphs in which all momenta are soft and the external 
momenta potential or ultrasoft. 
The graph is Taylor-expanded in the external ultrasoft momenta and in 
the zero components of external potential momenta. But it cannot 
be expanded in the spatial components of potential external momenta, 
because they are not small compared to the spatial components of the 
loop momenta. Hence the result is non-polynomial in the spatial 
components of the potential momenta. This is an instantaneous, but 
spatially non-local interaction. We call such interactions 
potentials. Potential gluons have propagators with $l_0^2$ expanded. 
They also give rise to potentials.

PNRQCD contains only potential quark fields and ultrasoft gluon (massless 
quark, ghost) fields. NRQCD is matched on PNRQCD order by order in 
$\alpha_s$. Consider quark-antiquark scattering at small relative 
momentum. At leading order in $\alpha_s$ the quark and antiquark 
interact by the exchange of a potential gluon. The leading term 
in $v$ yields the Coulomb potential at order $\alpha_s$. The 
corresponding non-local operator is
\begin{equation}
\label{nl}
\int d^3\vec{r}\,\left[\psi^\dagger T^A\psi\right](x+\vec{r}\,)\,
\left(-\frac{\alpha_s}{r}\right)\left[\chi^\dagger T^A\chi\right](x).
\end{equation}
The corrections of order $v^2$ are known as the Breit potential. At 
order $\alpha_s^2$ one has to compute the soft and potential 
contributions to the 1-loop NRQCD graphs and subtract the PNRQCD 
graphs constructed from the order-$\alpha_s$ potentials in the 
PNRQCD Lagrangian. The soft contributions to NRQCD graphs have no 
analogue in PNRQCD and renormalize the PNRQCD interactions. The 1-loop 
correction to the Coulomb potential is generated in this way 
together with a potential of form $\alpha_s^2/(m r^2)$ and higher 
order terms in $v$. For the potential contributions 
to NRQCD graphs it is necessary to perform an explicit matching to 
avoid double counting. As mentioned above, the contributions from 
soft and potential gluons may look different in different gauges. 
But their sum and hence the PNRQCD Lagrangian is gauge-invariant. 

In general the PNRQCD Lagrangian can be written as
\begin{equation}
{\cal L}_{\rm PNRQCD}={\cal L}^\prime_{\rm NRQCD} + 
{\cal L}_{\rm non-local},
\end{equation}
where ${\cal L}_{\rm non-local}$ collects all non-local interactions. 
The local interactions are exactly those of NRQCD, but the interpretation 
is different, because only ultrasoft gluons are left over. 
In loop graphs constructed from ${\cal L}^\prime_{\rm NRQCD}$ the 
gluon propagators are always expanded according to their ultrasoft 
scaling rule, while in loop graphs constructed from ${\cal L}_{\rm NRQCD}$ 
gluons are also soft and potential. The prime reminds us of this 
difference. 

Because only potential quarks and ultrasoft gluons are left in PNRQCD, 
the interaction terms have definite scaling rules. They agree with 
those given in Refs.~\citer{LAB,LS}. Note that the NRQCD scaling rules 
of Ref.~\cite{LMM} are really those of PNRQCD.
A potential quark propagator in coordinate space scales as $v^3$, so 
a quark field in PNRQCD scales as $v^{3/2}$. Comparing the scaling of 
$\psi^\dagger\partial^0\psi$ with the scaling of (\ref{nl}), 
we find that the former scales as $v^5$ and the latter scales 
as $\alpha_s v^4$. As is well-known, the Coulomb interaction 
cannot be treated as a perturbation when $v\sim \alpha_s(m v)$. 
Note, however, that the matching on PNRQCD can be done by treating 
the Coulomb interaction as a perturbation. The unperturbed PNRQCD 
Lagrangian is 
\begin{eqnarray}
\label{pnrqcd}
{\cal L}_{\rm PNRQCD}^0 &=& \psi^\dagger \left(i \partial^0+
\frac{\vec{\partial}^2}{2 m}
\right)\psi + 
\chi^\dagger \left(i \partial^0-\frac{\vec{\partial}^2}{2 m}
\right)\chi 
\nonumber\\
&&\hspace*{-1.5cm}
+\int d^3\vec{r}\,\left[\psi^\dagger T^A\psi\right](\vec{r}\,)\,
\left(-\frac{\alpha_s}{r}\right)\left[\chi^\dagger T^A\chi\right](0) +
{\cal L}_{\rm light}^{\rm free}.
\end{eqnarray}
One can rewrite this in terms of a `tensor field' $[\psi\otimes
\chi^\dagger](\vec{R},\vec{r}\,)$ that depends on the cms and 
relative coordinates. The unperturbed Lagrangian describes free 
propagation (with mass $2 m$) in the cms coordinate. The propagation 
of $[\psi\otimes \chi^\dagger](\vec{R},\vec{r}\,)$ in its 
relative coordinate is given by the Coulomb Green function of a 
particle with reduced mass $m/2$. In calculating diagrams with Coulomb 
Green functions, one sums corrections of order $(\alpha_s/v)^n$ to 
all orders. The remaining terms can be treated as perturbations in 
$v$ and $\alpha_s$ around the unperturbed Lagrangian. These are 
calculations familiar from QED bound state problems. What is new, 
in our opinion, is that we understand how to perform such calculations 
systematically in dimensional regularization, without double counting, 
and, if necessary, including retardation effects (graphs with 
ultrasoft lines). 

When an ultrasoft gluon line with momentum $l$ connects to a 
quark line with loop momentum $k-l/2$ for the incoming and $k+l/2$
for the outgoing quark line, the threshold expansion 
instructs us to expand the quark-gluon vertex 
and quark propagator in $\vec{l}/{\vec{k}}\sim v$. All gluon interaction 
terms in ${\cal L}^\prime_{\rm NRQCD}$ should be understood as 
multipole-expanded, for instance
\begin{equation}
\left[\psi^\dagger\vec{A}\cdot\vec{\partial}\,
\psi\right](x) \equiv
\psi^\dagger(t,\vec{x}\,)\vec{A}(t,0)\cdot\vec{\partial}\,
\psi(t,\vec{x}\,) + 
\psi^\dagger(t,\vec{x}\,)\,(\vec{x}\cdot\vec{\partial})\vec{A}(t,0)
\cdot\vec{\partial}\, \psi(t,\vec{x}\,) + \ldots,
\end{equation}
and likewise for all other interactions. 

Up to this point we have neglected the fact that in QCD -- contrary 
to QED -- the coupling constant evolves below the scale $m$. 
When $m v^2\ll m$, but 
$\alpha_s(m v^2) \ll 1$, one can sum up logarithms of $v$, but 
otherwise the power counting remains unaffected. In particular, 
only the Coulomb interaction has to be treated non-perturbatively. 
When $m v^2\sim \Lambda_{\rm QCD}$ the situation changes, because 
ultrasoft gluons couple with unit strength, since 
$\alpha_s(m v^2)\sim 1$. The coupling to heavy quarks is still 
small, of order $v$ at least, but the self-coupling of gluons is 
unsuppressed. An ultrasoft gluon propagator in coordinate space 
scales as $v^4$, hence the gluon kinetic term scales as $v^8$. 
The three-gluon and four-gluon vertices also scale as $v^8$. 
When $\alpha_s(m v^2)\sim 1$, ultrasoft gluons enter the calculation 
of $R_{Q\bar{Q}}$ as a non-perturbative contribution of relative order 
$v^2$ and `retardation effects' cannot be neglected at NNLO. 
A perturbative treatment of the problem cannot be extended 
beyond NLO, because the unperturbed PNRQCD Lagrangian must 
be modified to contain ${\cal L}_{\rm light}$ rather than 
${\cal L}_{\rm light}^{\rm free}$. The unperturbed problem 
is then no longer exactly solvable. This is why the energy spectra of 
charmonia and bottomonia are not exactly Coulomb-like.

\section{$J/\psi\to l^+ l^-$}

The two-loop short-distance correction to the leptonic decay of 
an $S$-wave quarkonium state such as $J/\psi$, $\psi'$ and $\Upsilon(nS)$
has been analyzed in Ref.~\cite{BSS98}. The decay rate can 
be expressed as 
\begin{equation}
\Gamma(J/\psi\to l^+ l^-) = \left(\,\sum_{k=0} c_k(\mu) 
\left(\frac{\alpha_s(m_c)}{\pi}\right)^{\!k}\,\right)^{\!2} 
\frac{4\pi e_c^2\alpha_{em}^2}{3 M_{J/\psi}}\,
\frac{12\,|\Psi(0)|^2(\mu)}{M_{J/\psi}},
\end{equation}
neglecting relativistic corrections which can be added 
systematically \cite{BBL} at the expense 
of further non-perturbative parameters 
in addition to the wave-function at the 
origin. The coefficients $c_k$ are those of (\ref{current}) (up to 
a normalization)  and $\Psi(0)$ is related to the 
vacuum-to-$J/\psi$ matrix element of the non-relativistic current. 
We have $c_0=1$, $c_1=-2 C_F$ \cite{oneloop,old} and the 
2-loop coefficient in the $\overline{\rm MS}$ factorization scheme 
reads~\cite{BSS98,CM98}
\begin{eqnarray}
\label{c2}
c_2(\mu) &=& C_F^2 \left\{\,
\pi^2\left[\frac{1}{6}\,\ln\left(\frac{m_Q^2}{\mu^2}\right) -
\frac{79}{36}+\ln 2\right] + \frac{23}{8}-\frac{\zeta(3)}{2}\right\}
\nonumber\\
&&\hspace*{-1.5cm}
+\,C_F C_A \left\{\pi^2\left[\frac{1}{4}\,
\ln\left(\frac{m_Q^2}{\mu^2}\right) +
\frac{89}{144}-\frac{5}{6}\,\ln 2\right] - \frac{151}{72}-
\frac{13\zeta(3)}{4}\right\}
+ \frac{11}{18}\,C_F T_F\,n_f 
\nonumber\\
&&\hspace*{-1.5cm}
+ \,C_F T_F \left\{
-\frac{2\pi^2}{9}+\frac{22}{9}\right\}. 
\end{eqnarray}
The calculation amounts to picking up the hard contributions only 
of the 2-loop three-point integrals contributing to 
$c+\bar{c}\to \gamma^*$. Numerically
\begin{equation}
\sum_{k=0} c_k(\mu) 
\left(\frac{\alpha_s(m_c)}{\pi}\right)^k
= 1-\frac{8\alpha_s(m_c)}{3\pi} - \left[44.55-0.41 n_f -25.59\,
\ln\frac{m_c}{\mu}\right]\left(\frac{\alpha_s}{\pi}\right)^2
+\ldots.
\end{equation}
Take $\mu=m_c$ for the factorization scale. Before squaring the 
coefficient function, the 1-loop correction is $-25$\%, but 
the 2-loop correction amounts to $-50$\%. Even for bottomonium, the 
2-loop correction is as large as the 1-loop correction.

At two loops the wavefunction at the origin becomes factorization 
scale and scheme\--de\-pen\-dent. 
The anomalous dimension is very large. 
The leptonic width is an important observable with respect to 
tuning the parameters of potential models. The large scheme-dependence 
is a problem for potential models, because it is not clear which 
scheme the wavefunction at the origin in potential models 
corresponds to. 

The above result suggests that the perturbative expansion is not 
reliable, so that perturbative factorization would not work 
quantitatively. But the large coefficients could also be the 
consequence of a `bad' factorization scheme. We will know only once 
a second quarkonium decay such as $\eta_c\to\gamma\gamma$ is computed 
to second order. With present analytic methods this seems to be a 
challenging piece of work.

\section{$t\bar{t}$ Production}

\subsection{The total cross section}

The total cross section for $e^+ e^-\to t\bar{t}+X$ has been 
calculated at NNLO in Refs.~\cite{HT98,MY98} for the vector 
coupling of the $t\bar{t}$ pair. (The axial-vector contribution is a 
NNLO effect but has not been considered in Refs.~\cite{HT98,MY98}.) 
This can be done by combining the hard matching coefficient (\ref{c2}) 
with the integrals over Coulomb Green functions in dimensional 
regularization. In practice, Ref.~\cite{HT98,MY98} used a different 
factorization scheme by comparing the resummed cross section 
with the cross section at the threshold at fixed order $\alpha_s^2$ 
\cite{CM98}. The final result is independent of this choice. 

The top quark is unstable on a scale comparable to the scale 
$m_t\alpha_s^2$ and finite width effects are essential in the 
threshold region. Following Ref.~\cite{KK}, the finite 
top quark width is taken into account by the substitution
\begin{equation}
\label{width}
E=\sqrt{s}-2 m_t \to E+i\,\Gamma_t
\end{equation}
in the argument of the Coulomb Green function, 
where $E$ is the energy measured from the threshold, defined 
by twice the top quark pole mass. Note that $E$ depends on the 
renormalization convention for the top quark mass and is not a 
physical quantity, contrary to the cms energy $\sqrt{s}$. The 
prescription (\ref{width}) has been justified in Ref.~\cite{KK} at LO. It 
is probably not justified at NNLO. In general, one can expect 
that the finite width inhibits the radiation of ultrasoft gluons 
from top quarks but affects the potential interactions less 
strongly.

\begin{figure}[t]
   \vspace{-4cm}
   \epsfysize=15cm
   \epsfxsize=10cm
   \centerline{\epsffile{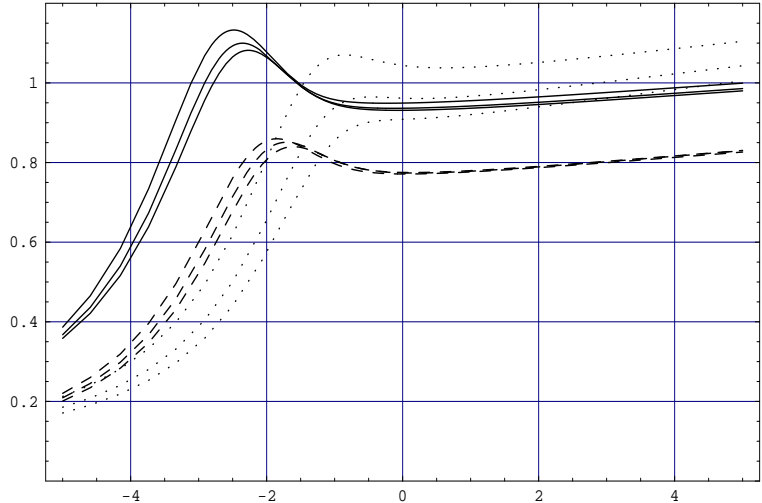}}
   \vspace*{-4cm}
\caption[dummy]{\small The $t\bar{t}$ cross section as function 
of $\sqrt{s}-2 m_t$ (in GeV) in LO (dotted), NLO (dashed) and 
NNLO (solid) for three choices of renormalization scales each. 
Parameters: $m_t=175\,$GeV, $\Gamma_t=1.43\,$GeV and $
\alpha_s(m_Z)=0.118$. Figure reprinted from Ref.~\cite{MY98}.
\label{fig1}}
\end{figure}
The result for the cross section in units of the point 
cross section as a function of $E$ is reproduced in Figure~\ref{fig1}. 
We note that the NNLO calculation has an as large effect on the 
height of the peak as the NLO calculation. Furthermore both 
shift the location of the peak by somewhat less than a GeV. 
If the QCD corrections do not converge, this implies an uncertainty 
in $m_t$ of almost $500\,$MeV, with frustrating consequences 
for precision studies of the $t\bar{t}$ threshold at a Next Linear 
Collider.

\subsection{Which mass?}

There is reason to believe that the situation is not as bad. When 
one discusses uncertainties in quark masses in the range of a few 
hundred MeV, it is important to ask how sensitive a particular 
mass renormalization prescription is to long-distance QCD effects. 
One has to be particularly careful about this, if the physical process 
is intrinsically less long-distance sensitive than the mass 
renormalization convention one is about to use.

For the pole mass of a stable quark long-distance sensitivity causes 
an uncertainty of order $\Lambda_{\rm QCD}$ \cite{BB94,BSUV94}. The 
analysis of Refs.~\cite{BB94,BSUV94} goes through unmodified 
\cite{SW97} for an unstable quark, when the pole mass is defined 
as the real part of the complex pole position in the top quark 
propagator. In perturbative calculations this long-distance 
sensitivity shows up as large radiative corrections, when the top 
quark pole mass as an input parameter is fixed as done above. It 
is advantageous to fix instead a mass renormalization scheme which 
is less sensitive to long-distances, provided one can show that the 
large corrections cancel then. (Effectively, this amounts to comparing 
a LO, NLO etc. calculation in terms of the pole mass at somewhat 
different input values of the pole mass. Because of the definition 
(\ref{width}) of $E$, this leads to a shift in the horizontal scale 
of Figure~\ref{fig1}, which compensates the shift in the LO, NLO 
and NNLO curves. It seems preferred to plot the cross section as a 
function of the physical parameter $\sqrt{s}$ rather than the 
scheme-dependent parameter $E$.)

Indeed the analysis of higher-order radiative corrections to the 
Coulomb potential reveals systematically large terms~\citer{AL95,JPS98}. 
In part, these large corrections are a consequence of long-distance 
sensitivity. The long-distance sensitive contributions to the 
Coulomb potential can be shown~\cite{B98,HSSW98} to cancel exactly 
against those in the pole mass renormalization. Contrary to 
intuition the threshold cross section is less sensitive to 
long distances than the pole mass and hence the threshold is not 
tied to twice the pole mass once we talk about 
accuracies of several hundred MeV. One can make this 
cancellation manifest~\cite{B98} by subtracting the dangerously 
large terms from the potential and by adding them back to the 
pole mass. The result is a modified mass renormalization prescription, 
called the `potential subtracted mass', which can be related to 
a more conventional definition like the $\overline{\rm MS}$ mass by 
a reasonably well-behaved series, but which is at the same time not 
unphysically 
far away from the threshold like the $\overline{\rm MS}$ mass (at 
the scale $m_t$) itself. It is of course possible to use other 
modified mass renormalization prescriptions, provided they also 
lead to a manifest cancellation of long-distance sensitive terms and 
can be computed to sufficient accuracy. One possible alternative 
is discussed in Ref.~\cite{CMU97}.

The discussion of the previous paragraph applies to a stable or 
unstable quark. Also, as mentioned, the top quark pole mass suffers 
from the same long-distance sensitivity as the bottom or charm 
quark pole mass despite the fact that the width of the top quark 
is significantly larger than $\Lambda_{\rm QCD}$. Still, the 
width helps. The point is not that the top quark pole mass should 
be better behaved. The point is that contrary to bottom quarks, 
where one can find observables (such as the $B$ meson mass) which 
are as long-distance sensitive as the pole mass, there is no 
observable involving top quarks that would be as sensitive to 
long distances as the top quark pole mass. In this precise sense, 
the top quark pole mass is an irrelevant quantity. The top quark 
propagator is simply never on-shell. The propagator is given by 
$1/(\not\!p-m_t+i\Gamma_t/2)$, but the loop or external momentum $p$ 
can always be considered as real. The denominator of the propagator 
never gets smaller than about $m_t\Gamma_t$. We then find, for a quantity 
that would have a long-distance correction of (relative) 
order $\Lambda_{\rm QCD}/m$ for a stable quark,
\begin{equation}
\frac{\Lambda_{\rm QCD}}{m} \to \frac{\Lambda_{\rm QCD}}{m} \cdot 
\frac{\Lambda_{\rm QCD}}{\Gamma}.
\end{equation}
We expect, at least, a suppression by $\Lambda_{\rm QCD}/\Gamma_t$ due 
to the finite width. (For a related discussion of this point, see 
Refs.~\cite{khoze,big}.)

\section{Bottom quark mass from Upsilon sum rules}

The derivatives of the bottom vector current two-point function at 
$q^2=0$ are 
related to the inclusive bottom cross section by 
\begin{equation}
\frac{12\pi^2}{n!}\,\frac{d^n}{d(q^2)^n}\,\Pi(q^2)_{\big| q^2=0} = 
\int_0^\infty \frac{ds}{s^{n+1}}\,R_{b\bar{b}}(s)
\end{equation}
up to a (small) correction due to $b\bar{b}$ radiation from light 
quarks. The left hand side can be computed in perturbation theory; 
the right hand side from data.

The parameters of the lowest $\Upsilon(nS)$ resonances are well-measured, 
but the continuum cross section near the threshold is not. Hence the 
experimental error of the right hand side decreases with increasing $n$, 
because higher moments weight lower $s$. But larger $n$ makes the 
theoretical calculation more difficult, because the expansion parameter 
is $\alpha_s\sqrt{n}$. The enhancement by a factor $\sqrt{n}$ is a 
consequence of the Coulomb enhanced terms in the theoretical calculation 
of $R$. A power of $v$ in $R$ corresponds to a power of $n^{-1/2}$ in 
the moment. When 
\begin{equation}
\sqrt{\pi n}\,\alpha_s(m_b/\sqrt{n}) \ll 1
\end{equation}
is no longer satisfied, a resummation by the methods discussed earlier 
in this talk is necessary. This is already the case for $n$ larger than 4. 
There is an intermediate $n$ where experimental and theoretical 
uncertainties are balanced.

\begin{table}[t]
\addtolength{\arraycolsep}{0.2cm}
\renewcommand{\arraystretch}{1.7}
$$
\begin{array}{|c|c|c|c|c|}
\hline
  & \mbox{Order} & n & m_b^{\rm pole} & 
m_b^{\overline{\rm MS}} \\ 
\hline
\mbox{Ref.~${}^{46}$}  
& \mbox{NLO}  & \mbox{8 - 20}  & 4.827\pm 0.007 & \mbox{--}    \\
\mbox{Ref.~${}^{47}$}  
& \alpha_s^2  & \mbox{8 - 20}  & 4.604\pm 0.014 & 4.13\pm 0.06 \\
\mbox{Ref.~${}^{48}$}  
& \mbox{NLO}  & \mbox{10 - 20} & 4.75\pm 0.04   & \mbox{--}    \\
\mbox{Ref.~${}^{28}$}  
& \mbox{NNLO} & \mbox{10 - 20} & 4.78\pm 0.04   & \mbox{--}    \\
\mbox{Ref.~${}^{29}$}  
& \mbox{NNLO} & \mbox{4 - 10 } & \mbox{4.78 - 4.98} & \mbox{4.16 - 4.33} \\
\mbox{Ref.~${}^{30}$}  
& \mbox{NNLO} & \mbox{14 - 18} & \mbox{--}      & 4.20\pm 0.1  \\
\hline
\end{array}
$$
\caption[dummy]{\label{tab1}\small 
Bottom quark mass obtained from $\Upsilon$ sum rules in GeV. The 
$\overline{\rm MS}$ mass is renormalized at the scale of its own value. The 
different values are not strictly comparable, because different values 
of $\alpha_s(m_Z)$ may be implied. Whenever available, we quote a result 
obtained from a fixed input rather than fitted $\alpha_s(m_Z)$. 
`$\alpha_s^2$' means that no systematic resummation has been attempted.}
\end{table}
Recently, there have been several calculations~\citer{PP98,MY98II} which 
implemented this resummation at NNLO. Their results, together with 
results of previous (NLO or fixed-order $\alpha_s^2$) 
calculations~\citer{VOL95,KKP98} are compiled in Table~\ref{tab1}. 
Note that as the accuracy of the calculation increases, the uncertainty 
in the result becomes larger. Instead of our own comments, we refer to 
Ref.~\cite{Hoa98} for a discussion of this intriguing point. The result 
of Ref.~\cite{JP95} is low, mainly because they do not include the 
sub-threshold poles of the Coulomb Green function into their 
calculation. But since to a large extent it is the average over 
Coulomb poles which is dual to the average of over the physical 
$\Upsilon$ resonances, their inclusion is necessary for moments which 
receive their largest contribution from the $\Upsilon$ resonances.
Technically, this follows from the fact that the Coulomb poles 
contribute and amount of order $\alpha_s(m_b/\sqrt{n})^3$ to the 
moments, to be compared with $1/n^{3/2}$, the tree graph contribution.

Note that because of the cancellation of long-distance contributions 
between the potential and the pole mass discussed in the previous 
section, it is advantageous to use a mass renormalization convention 
different from the pole mass beyond NLO. This strategy has been 
adopted in Refs.~\cite{MY98II,BSS98II}.

We mentioned earlier that there is a non-perturbative contribution 
to the heavy quark cross section at NNLO 
when $m v^2\sim \Lambda_{\rm QCD}$. There is a contribution to the 
moments from the scale $m_b/n$. If we require for a perturbative 
treatment that all scales are larger than $0.5\,$GeV, this limits 
$n<10$. The use of larger $n$ in most of the calculations quoted 
in Table~\ref{tab1} has been justified by the observation that 
even at $n=20$ the gluon condensate contribution to the moments 
is very small. Hence one may be in the fortunate situation that 
the actual non-perturbative contribution is smaller than the power 
counting argument suggests. However, the 
condensate expansion converges parametrically only when 
$m_b/n\gg \Lambda_{\rm QCD}$. If the expansion does not converge, 
the size of its first term is not very conclusive. (For 
the Coulomb energy levels, the contribution from dimension-6 operators 
has been considered in Ref.~\cite{Pin97}.)

\section{Conclusion}

The past year has seen significant progress in our understanding 
of perturbative quark-antiquark systems close to threshold, both 
methodical and in terms of explicit higher-order calculations. We have 
come nearer to the answer to the question how accurately the bottom and 
the top quark mass can be determined by purely perturbative means.

\section*{Acknowledgements}
I thank V.A.~Smirnov and A.~Signer for their collaboration on topics 
related to this work and P.~Labelle and V.A.~Smirnov 
for reading the manuscript. I thank the EU for financial support 
to attend the conference.


\section*{References}
\begin{thebibliography}{99}

\bibitem{NSVZ1}
V.A.~Novikov {\em et al.}, \Journal{\PRL}{38}{626}{1977} [Erratum: 
{\em ibid.} {\bf 38}, 791 (1977)].
\bibitem{NSVZ2}
V.A.~Novikov {\em et al.}, Phys. Rep. {\bf 41}, 1 (1978).
\bibitem{VZ87}
M.B.~Voloshin and Yu.M.~Zaitsev, Usp. Fiz. Nauk. {\bf 152}, 361 (1987) 
[Sov. Phys. Usp. {\bf 30}(7), 553 (1987)].
\bibitem{B86}
I.I.~Bigi {\em et al.}, \Journal{\PLB}{181}{157}{1986}.
\bibitem{KK}
V.S.~Fadin and V.A.~Khoze, Pis'ma Zh. Eksp. Teor. Fiz. {\bf 46}, 417 
(1987) [JETP Lett. {\bf 46}, 525 (1987)]; Yad. Fiz. {\bf 48}, 487 
(1988) [Sov. J. Nucl. Phys. {\bf 48}(2), 309 (1988)].
\bibitem{SP91}
M.J.~Strassler and M.E.~Peskin, \Journal{\PRD}{43}{1500}{1991}
\bibitem{BBL}
G.T.~Bodwin, E.~Braaten and G.P.~Lepage, \Journal{\PRD}{51}{1125}{1995} 
[Erratum: {\em ibid.} {\bf D55}, 5853 (1997)].
\bibitem{KS96}
N.~Kideonakis and G.~Sterman, \Journal{\PLB}{387}{867}{1996}; 
\Journal{\NPB}{505}{321}{1997}.
\bibitem{BCMN98}
R.~Bonciani, S.~Catani, M.L.~Mangano and P.~Nason, 
CERN-TH-98-31 [hep-ph/9801375].

\bibitem{LAB}
P.~Labelle, MCGILL-96-33 [hep-ph/9608491].
\bibitem{LM}
M.~Luke and A.V.~Manohar, \Journal{\PRD}{55}{4129}{1997}.
\bibitem{GR}
B.~Grinstein and I.Z.~Rothstein, \Journal{\PRD}{57}{78}{1998}.
\bibitem{LS}
M.~Luke and M.J.~Savage, \Journal{\PRD}{57}{413}{1998}.
\bibitem{PS}
A.~Pineda and J.~Soto, Nucl. Phys. Proc. Suppl. {\bf 64}, 428 (1998) 
[hep-ph/9707481].
\bibitem{BS}
M.~Beneke and V.A.~Smirnov, CERN-TH-97-315, to appear in Nucl. Phys.~B 
[hep-ph/9711391].
\bibitem{PS2}
A.~Pineda and J.~Soto, \Journal{\PLB}{420}{391}{1998}; 
UB-ECM-PF-98-11 [hep-ph/9805424]
\bibitem{Griesshammer}
H.W.~Grie\ss hammer, NT-UW-98-3 [hep-ph/9712467]; 
NT-UW98-12 [hep-ph/9804251].

\bibitem{CL86} 
W.E.~Caswell and G.P.~Lepage, \Journal{\PLB}{167}{437}{1986}.
\bibitem{TL}
B.A.~Thacker and G.P.~Lepage, \Journal{\PRD}{43}{196}{1991}.
\bibitem{LMM}
G.P.~Lepage {\em et al.}, 
\Journal{\PRD}{46}{4052}{1992}.

\bibitem{Pet97}
M.~Peter, \Journal{\PRL}{78}{602}{1997}; 
\Journal{\NPB}{501}{471}{1997}.
\bibitem{hoang}
A.H.~Hoang, \Journal{\PRD}{56}{7276}{1997}.
\bibitem{BSS98}
M.~Beneke, A.~Signer and V.A.~Smirnov, 
\Journal{\PRL}{80}{2535}{1998}.
\bibitem{CM98}
A.~Czarnecki and K.~Melnikov, \Journal{\PRL}{80}{2531}{1998}.

\bibitem{PY97}
A.~Pineda and F.J.~Yndurain, UB-ECM-PF-97-34 [hep-ph/9711287]
\bibitem{HT98}
A.H.~Hoang and T.~Teubner, UCSD/PTH 98-01 [hep-ph/9801397].
\bibitem{MY98}
K.~Melnikov and A.~Yelkhovsky, BudkerINP-98-7 [hep-ph/9802379].
\bibitem{PP98}
A.A.~Penin and A.A.~Pivovarov, TTP/98-13 [hep-ph/9803363].
\bibitem{Hoa98}
A.~Hoang, UCSD/PTH 98-02 [hep-ph/9803454].
\bibitem{MY98II}
K.~Melnikov and A.~Yelkhovsky, TTP98-17 [hep-ph/9805270].
\bibitem{BSS98II}
M.~Beneke, A.~Signer and V.A.~Smirnov, in preparation.

\bibitem{Ma97}
A.V.~Manohar, \Journal{\PRD}{56}{230}{1997}.
\bibitem{oneloop}
R.~Barbieri, R.~Gatto, R.~K\"orgerler and Z. Kunszt, 
\Journal{\PLB}{57}{455}{1975}.
\bibitem{old}
R.~Karplus and A.~Klein, Phys. Rev. {\bf 87}, 848 (1952).


\bibitem{BB94}
M.~Beneke and V.M.~Braun, \Journal{\NPB}{426}{301}{1994};
M.~Beneke, \Journal{\PLB}{344}{341}{1995}.
\bibitem{BSUV94}
I.I.~Bigi, M.A.~Shifman, N.G.~Uraltsev and A.I.~Vainshtein, 
\Journal{\PRD}{50}{2234}{1994}.
\bibitem{SW97}
M.C.~Smith and S.S.~Willenbrock, \Journal{\PRL}{79}{3825}{1997}.
\bibitem{AL95}
U.~Aglietti and Z.~Ligeti, \Journal{\PLB}{364}{75}{1995}.
\bibitem{JKPST98}
M.~Jezabek {\em et al.}, DESY-98-019 [hep-ph/9802373].
\bibitem{JPS98}
M.~Jezabek, M.~Peter and Y.~Sumino, HD-THEP-98-10 [hep-ph/9803337].
\bibitem{B98}
M.~Beneke, CERN-TH/98-120, to appear in Phys. Lett. B  
[hep-ph/9804241]. 
\bibitem{HSSW98}
A.H.~Hoang, M.C.~Smith, T.~Stelzer and S.~Willenbrock, UCSD/PTH 98-13 
[hep-ph/9804227].
\bibitem{CMU97}
A.~Czarnecki, K.~Melnikov and N.~Uraltsev, \Journal{\PRL}{80}{3189}{1998}.
\bibitem{khoze}
V.A.~Khoze, in: Proceedings of the 1st Arctic Workshop on 
Future Physics and Accelerators, Saariselka, Finland, 1994, 
(World Scientific, Singapore, 1995) [hep-ph/9411239].
\bibitem{big}
I. Bigi, M. Shifman and N. Uraltsev, Ann. Rev. Nucl. Part. Sci. {\bf47},
591 (1997) 

\bibitem{VOL95} 
M.B.~Voloshin, Int. J. Mod. Phys. {\bf A10}, 2865 (1995).
\bibitem{JP95}
M.~Jamin and A.~Pich, \Journal{\NPB}{507}{334}{1997}.
\bibitem{KKP98}
J.H.~K\"uhn, A.A.~Penin and A.A.~Pivovarov, TTP98-01 [hep-ph/9801356].
\bibitem{Pin97}
A.~Pineda, \Journal{\NPB}{494}{213}{1997}.


\end{thebibliography}
\end{document}